\documentclass[aps,prd,preprint,showpacs,showkeys,preprintnumbers,amsmath,amssymb]{revtex4-1}
\usepackage{amsmath}
\usepackage{amssymb}

\begin{document}

\title{On Gravitational Entropy of de Sitter Universe}
\author{S. C. Ulhoa }
\email{sc.ulhoa@gmail.com} \affiliation{Instituto de F\'{i}sica,
Universidade de Bras\'{i}lia, 70910-900, Bras\'{i}lia, DF,
Brazil.\\Faculdade Gama, Universidade de Bras\'{i}lia, Setor Leste
(Gama), 72444-240, Bras\'{i}lia-DF, Brazil.}

\author{E. P. Spaniol}
\email{spaniol.ep@gmail.com} \affiliation{UDF Centro
Universit\'ario and Centro Universit\'ario IESB (Campus Sul – Edson Machado), Bras\'ilia, DF, Brazil.}
\date{\today}

\begin{abstract}
The paper deals with the calculation of the gravitational entropy in
the context of teleparallel gravity for de Sitter space-time. In
such a theory it is possible to define gravitational energy and
pressure, thus we use those expressions to construct the
gravitational entropy. We interpret the cosmological constant as the
temperature and write the first law of thermodynamics. In the limit
$\Lambda\ll 1$ we find that the entropy is proportional to volume
and $\Delta S\geq 0$.
\end{abstract}



\maketitle
\section{Introduction}
\noindent

The idea of black hole thermodynamics started with the pioneering
works of Bekenstein and
Hawking~\cite{PhysRevD.7.2333,PhysRevD.13.191}. It was noted that
the area of the event horizon behaves as an entropy. Together with
such discovery it was also noted that it has a specific temperature.
Thus it radiates and evaporates leading to a lost on the information
inside the black hole, which apparently violates the second law of
thermodynamics. It was the so called information
paradox~\cite{PhysRevD.72.084013}. Dolan has pointed out that such
study was incomplete without the term $pdV$ in the first law of
thermodynamics~\cite{0264-9381-28-23-235017}. However the concept of
gravitational pressure is difficult to establish as the very idea of
gravitational energy. The matter of the definition of gravitational
energy has a long story and yet it is a very controversial theme.
The main approaches in this subject are Komar
integrals~\cite{Komar}, ADM formalism~\cite{ADM} and quasi-local
expressions~\cite{York,Sza1}. In opposition to general relativity,
in teleparrallel gravity those quantities can be well defined.

Teleparallel Equivalent to General Gelativity (TEGR) is an
alternative theory of gravitation constructed out in terms of the
tetrad field on Weitzenb\"ock geometry. It was first proposed by
Einstein in an attempt to derive an unified field
theory~\cite{einstein}. Later it was revived with a paper entitled
``New General Relativity''~\cite{Shirafuji}, since then a lot of
improvement have been made in the understanding of gravitational
energy and the role of torsion~\cite{Hehl, Hehl2}. In the context of
TEGR it is possible to define an expression for gravitational energy
which is invariant under coordinates transformation and dependent on
the reference frame. Those features are present in the special
theory of relativity and there is no physical reason to abandon such
ideas once dealing with a gravitational theory. Using the field
equations of TEGR, it is possible to define an expression for the
gravitational pressure. Therefore a natural extension is to define
an expression for the gravitational entropy. The advantage of this
procedure is define an entropy in terms of of purely thermodynamical
quantities such as energy and pressure. This will be our main goal
in this paper, for de Sitter Universe. This Universe model is
important because it describes an expanding empty space. Thus it is
possible to shed light into the vacuum energy and cosmological
inflationary models.

The paper is organized as follows. In Section \ref{tel}, we present
the main ideas of Teleparallel gravity. From field equation we
derive the total energy and pressure. In Section \ref{deSitter}, we
calculate such quantities for de Sitter Universe, then we use the
first law of Thermodynamics to get the gravitational entropy. To
achieve such aim, we have interpreted the cosmological constant as
the temperature of the system. Finally we present our concluding
remarks in the last section.

\bigskip
Notation: space-time indices $\mu, \nu, ...$ and SO(3,1) indices $a,
b, ...$ run from 0 to 3. Time and space indices are indicated
according to $\mu=0,i,\;\;a=(0),(i)$. The tetrad field is denoted by
$e^a\,_\mu$ and the determinant of the tetrad field is represented
by $e=\det(e^a\,_\mu)$. In addition we adopt units where
$G=k_b=c=1$, unless otherwise stated.\par
\bigskip

\section{Teleparallel
Equivalent to General Gelativity (TEGR)}\label{tel} \noindent

Teleparallel gravity is a theory entirely equivalent to general
relativity, however it is formulated in the framework of
Weitzenb\"ock geometry rather than in terms of Riemann geometry.
Weitzenb\"ock geometry is endowed with the Cartan
connection~\cite{Cartan}, given by
$\Gamma_{\mu\lambda\nu}=e^{a}\,_{\mu}\partial_{\lambda}e_{a\nu}$,
where $e^{a}\,_{\mu}$ is the tetrad field, thus the torsion tensor
can be calculated in terms of this field by
\begin{equation}
T^{a}\,_{\lambda\nu}=\partial_{\lambda} e^{a}\,_{\nu}-\partial_{\nu}
e^{a}\,_{\lambda}\,. \label{3}
\end{equation}
Such a geometry keeps a relation to a Riemannian manifold, for
instance we note the the Cartan connection and Christofell
(${}^0\Gamma_{\mu \lambda\nu}$) symbols are related by a
mathematical identity, which reads

\begin{equation}
\Gamma_{\mu \lambda\nu}= {}^0\Gamma_{\mu \lambda\nu}+ K_{\mu
\lambda\nu}\,, \label{2}
\end{equation}
where

\begin{eqnarray}
K_{\mu\lambda\nu}&=&\frac{1}{2}(T_{\lambda\mu\nu}+T_{\nu\lambda\mu}+T_{\mu\lambda\nu})\label{3.5}
\end{eqnarray}
is the contortion tensor. The tetrad field, which is the dynamical
variable of the theory, is obtained from the metric tensor. Thus
Weitzenb\"ock geometry is less restrictive than Riemann geometry,
for each metric tensor it is possible to construct a infinite number
of tetrad fields. This apparent arbitrary behavior is amended once
we recall the interpretation of the tetrad field. The component
$e_{(0)}\,^\mu$ is associated to the four-velocity of the observer,
then for each reference frame there exists only one tetrad field. In
fact we have to make use of the acceleration tensor to completely
settle the state of an observer~\cite{Maluf:2007qq}, since it could
be in rotation as well in translation.

If one tries to construct the curvature from the Cartan connection,
he/she will find out that it vanishes identically. Hence the
Weitzenb\"ock geometry is described by a vanishing curvature and the
presence of torsion. The Riemann geometry, as it is well known, has
a vanishing torsion and a non-vanishing curvature tensor. Therefore
making use of the identity (\ref{2}) to construct the scalar
curvature, it leads to

\begin{equation}
eR(e)\equiv -e({1\over 4}T^{abc}T_{abc}+{1\over
2}T^{abc}T_{bac}-T^aT_a)+2\partial_\mu(eT^\mu)\,,\label{5}
\end{equation}
where $e$ is the determinant of the tetrad field and $R(e)$ is the
scalar curvature constructed out in terms of such a field. It should
be noted that the metric tensor alone does not establish a geometry.
From the above identity, we see that it is possible to construct a
tetrad field adapted to a specific reference frame which induces,
for the same metric tensor, both, a curvature in riemannian manifold
and torsion in a Weitzenb\"ock geometry. By means the very same
above identity, it is possible to find the counterpart of
Hilbert-Einstein Lagrangian density for teleparallel gravity, thus,
up to a total divergence (which plays no role in the field
equations), it reads

\begin{equation}
\mathfrak{L}= -k e({1\over 4}T^{abc}T_{abc}+{1\over
2}T^{abc}T_{bac}- T^aT_a) -\mathfrak{L}_M \,,
\end{equation}
where $k=1/16\pi$, $T_a=T^b\,_{ba}$, $T_{abc}=e_b\,^\mu e_c\,^\nu
T_{a\mu\nu}$ and $\mathfrak{L}_M$ stands for the Lagrangian density
for the matter fields. This Lagrangian density can be rewritten as

\begin{equation}
\mathfrak{L}\equiv -ke\Sigma^{abc}T_{abc} -\mathfrak{L}_M\,,
\label{5}
\end{equation}
where

\begin{equation}
\Sigma^{abc}={1\over 4} (T^{abc}+T^{bac}-T^{cab}) +{1\over 2}(
\eta^{ac}T^b-\eta^{ab}T^c)\,. \label{6}
\end{equation}

The field equations can be derived from Lagrangian (\ref{5}) using a
variational derivative with respect to $e^{a\mu}$, they read

\begin{equation}
e_{a\lambda}e_{b\mu}\partial_\nu (e\Sigma^{b\lambda \nu} )- e
(\Sigma^{b\nu}\,_aT_{b\nu\mu}- {1\over 4}e_{a\mu}T_{bcd}\Sigma^{bcd}
)={1\over {4k}}eT_{a\mu}\,, \label{7}
\end{equation}
where $\delta \mathfrak{L}_M / \delta e^{a\mu}=eT_{a\mu}$. The field
equations may be rewritten as

\begin{equation}
\partial_\nu(e\Sigma^{a\lambda\nu})={1\over {4k}}
e\, e^a\,_\mu( t^{\lambda \mu} + T^{\lambda \mu})\;, \label{8}
\end{equation}
where $T^{\lambda\mu}=e_a\,^{\lambda}T^{a\mu}$ and

\begin{equation}
t^{\lambda \mu}=k(4\Sigma^{bc\lambda}T_{bc}\,^\mu- g^{\lambda
\mu}\Sigma^{bcd}T_{bcd})\,. \label{9}
\end{equation}
In view of the antisymmetry property
$\Sigma^{a\mu\nu}=-\Sigma^{a\nu\mu}$, it follows that

\begin{equation}
\partial_\lambda
\left[e\, e^a\,_\mu( t^{\lambda \mu} + T^{\lambda \mu})\right]=0\,.
\label{10}
\end{equation}
Such equation leads to the following continuity equation,

\begin{equation}
{d\over {dt}} \int_V d^3x\,e\,e^a\,_\mu (t^{0\mu} +T^{0\mu})
=-\oint_S dS_j\, \left[e\,e^a\,_\mu (t^{j\mu} +T^{j\mu})\right]\,.
\label{11}
\end{equation}
Therefore we identify $t^{\lambda\mu}$ as the gravitational
energy-momentum tensor~\cite{PhysRevLett.84.4533,maluf2}.

Then, as usual, the total energy-momentum vector is defined
by~\cite{Maluf:2002zc}

\begin{equation}
P^a=\int_V d^3x\,e\,e^a\,_\mu (t^{0\mu} +T^{0\mu})\,, \label{12}
\end{equation}
where $V$ is a volume of the three-dimensional space. It is
important to note that the above expression is invariant under
coordinate transformations and it transforms like a 4-vector under
Lorentz transformations. The energy-momentum flux is given by the
time derivative of (\ref{12}), thus by means (\ref{11}) we find

\begin{equation}
\Phi^a=\oint_S dS_j\, \, e\,e^a\,_\mu \left(t^{j\mu}+
T^{j\mu}\right)\,. \label{13}
\end{equation}

If we assume a vacuum solution, e.g. a vanishing energy-momentum
tensor of matter fields, then we have

\begin{equation}
{{dP^a}\over {dt}}= -\oint_S dS_j\, \left[e\,e^a\,_\mu t^{j\mu}
\right]\,. \label{16}
\end{equation}
which is the gravitational energy-momentum flux~\cite{MF}. Using the
field equations (\ref{8}), the total energy-momentum flux reads

\begin{equation}
{{dP^a}\over {dt}}= -4k\oint_S dS_j\,
\partial_\nu(e\Sigma^{a j\nu})\,.
\label{17}
\end{equation}
Now let us restrict our attention to the spatial part of the
energy-momentum flux, i.e. the momentum flux, we have

\begin{equation}
{{dP^{(i)}}\over {dt}}= -\oint_S dS_j\, \phi^{(i)j} \,, \label{18}
\end{equation}
where

\begin{equation}
\phi^{(i)j}=4k\partial_\nu(e\Sigma^{(i)j\nu}) \,, \label{19}
\end{equation}
we note that the momentum flux is precisely the force, hence, since
 $dS_j$ is an element of area, we see that
$-\phi^{(i)j}$ represents the pressure along the $(i)$ direction,
over and element of area oriented along the $j$
direction~\cite{maluf2}. It should be noted that all definitions
presented in this section follow exclusively from the field
equations (\ref{8}).

\section{The Gravitational Entropy for de Sitter Spacetime}\label{deSitter}
\noindent

The de Sitter space-time is defined by the following line element

\begin{equation}
ds^2=-\biggl( 1-{{r^2}\over{R^2}}\biggr)dt^2+ \biggl(
1-{{r^2}\over{R^2}}\biggr)^{-1}dr^2+ r^2d\theta^2+r^2\sin^2\theta
d\phi^2\;,
\end{equation}
where $R=\sqrt{3\over \Lambda}$ and $\Lambda$ is the cosmological
constant. Such a space-time works as a model of an expanding
Universe, thus many inflationary models of the early Universe make
use of this feature~\cite{Nambu,Seery:2009hs,Lasenby:2003ur}.

Let us choose the following tetrad field adapted to a stationary
reference frame
\begin{equation}
e^a\,_\mu= \left[ \begin {array}{cccc} A&0&0&0
\\\noalign{\medskip}0&A^{-1}\sin\theta\cos\phi &r\cos\theta\cos\phi&-r\sin\theta\sin\phi
\\\noalign{\medskip}0&A^{-1}\sin\theta\sin\phi&r\cos\theta\sin\phi
&r\sin\theta\cos\phi\\\noalign{\medskip}0&A^{-1}\cos\theta
&-r\sin\theta&0\end {array} \right]
\end{equation}
where $A=\biggl( 1-{{r^2}\over{R^2}}\biggr)^{{1\over 2}}$ and its
determinant is $e=r^2\sin\theta$. Thus the non-vanishing components
of the torsion tensor are

\begin{eqnarray}
T_{001}&=&-\frac{r}{R^2}\,,\nonumber\\
T_{212}&=&r \Big[ 1- \big(1-\frac{r^2}{R^2} \big)^{- 1/2} \Big]\,,\nonumber\\
T_{313}&=&r \sin^2\theta \Big[ 1- \big(1-\frac{r^2}{R^2} \big)^{-
1/2} \Big] \,.
\end{eqnarray}
In order to calculate the gravitational energy and pressure we need
the components $\Sigma^{\mu\nu\lambda}$, thus after some algebraic
manipulation we find that the non-vanishing ones are

\begin{eqnarray}
\Sigma^{001}&=&\frac{1}{r}  \left[ \left( 1- \frac{r^2}{R^2}
\right)^{-\frac{1}{2}} -1 \right]\,,\nonumber\\
\Sigma^{212}&=&\frac{1}{2r^3} \left[ \sqrt{1-\frac{r^2}{R^2}}-
\left(1-\frac{2r^2}{R^2} \right) \right]\,,\nonumber\\
\Sigma^{313}&=&\frac{1}{2r^3 \sin^2\theta} \left[
\sqrt{1-\frac{r^2}{R^2}}- \left(1-\frac{2r^2}{R^2} \right)
\right]\,.\label{Sigma}
\end{eqnarray}
Hence, substituting the above components into  eq. (\ref{12}) for
$a=(0)$, it is possible to find the total energy and it reads

\begin{equation}
E=r_0\left(1-\sqrt{1-\frac{r_0^2}{R^2}}\right)\,,
\end{equation}
where $r_0$ is the radius of a spherical 3-dimensional hypersurface
of integration. In addition we have make the identification $E\equiv
P^{(0)}$. It should be noted that such expression already appeared
on reference~\cite{Ulhoa:2010wv}.

Similarly the radial pressure can be constructed from the components
$\phi^{(i)j}$ which is given in terms of the components in
(\ref{Sigma}). Thus after some simple calculations, we find

\begin{eqnarray}
\phi^{(1)1}&=& - 4k \cos\phi \sin^2\theta \left[
\sqrt{1-\frac{r^2}{R^2}}- \left(1-\frac{2r^2}{R^2} \right)
\right]\,,\nonumber\\
\phi^{(2)1}&=& - 4k \sin\phi \sin^2\theta \left[
\sqrt{1-\frac{r^2}{R^2}}- \left(1-\frac{2r^2}{R^2} \right)
\right]\,,\nonumber\\
\phi^{(3)1}&=& - 4k \sin \theta \cos\theta \left[
\sqrt{1-\frac{r^2}{R^2}}- \left(1-\frac{2r^2}{R^2} \right)
\right]\,.
\end{eqnarray}
Then we construct a radial $\phi$, which is denoted by
$\phi^{(r)1}$, by means the relation

\begin{eqnarray}
\phi^{(r)1}= \sin \theta \cos\phi \phi^{(1)1} + \sin \theta
\sin\phi\phi^{(2)1} + \cos\theta \phi^{(3)1} \,,
\end{eqnarray}
it yields

\begin{eqnarray}
\phi^{(r)1}= - 4k \sin \theta \left[ \sqrt{1-\frac{r^2}{R^2}}-
\left(1-\frac{2r^2}{R^2} \right) \right] \,.
\end{eqnarray}
The radial pressure is given by

\begin{eqnarray}
p(r)= \int_{0}^{2\pi} d \phi \int_{0}^{\pi} d \theta (-\phi^{(r)1})
\,,
\end{eqnarray}
therefore, we obtain

\begin{eqnarray}
p(r)= \left[ \sqrt{1-\frac{r^2}{R^2}}- \left(1-\frac{2r^2}{R^2}
\right) \right] \ \,.
\end{eqnarray}

Once the energy and pressure are given, we can turn out our
attention to the entropy itself. It is a thermodynamical potential
linked to the variation of energy and volume, i.e.

\begin{equation}
TdS=dE+pdV\,,
\end{equation}
where T is the temperature (considered constant) and S is the
entropy. The change in the volume is realized through the variation
of $r_0$, thus $pdV=p(r_0)dr_0$. The total energy also changes with
the variation of $r_0$. In this system we have two parameters $r_0$
and the cosmological constant, the first one is linked to the volume
while the second one has to play the role of a temperature. As a
consequence we suppose the temperature could be given by an
arbitrary function of the cosmological constat, $T=f(\Lambda)$.
Therefore the energy will vary with the change of the temperature.
We'll use the cosmological constant as the temperature. Therefore
the first law of Thermodynamics simply reads

\begin{equation}
\Lambda dS=\frac{\partial E}{\partial r_0}\,dr_0+p(r_0)\,dr_0\,,
\end{equation}
then the gravitational entropy is given by

\begin{equation*}
S(r_0)=\left(\frac{1}{\Lambda }\right)\int\left(\frac{\partial
E}{\partial r_0}+p(r_0)\right)\,dr_0\,,
\end{equation*}
which yields

\begin{equation}
S(r_0)=\left(\frac{1}{\Lambda
}\right)\left[r_0\left(\frac{2r_0^2}{3R^2}-\frac{1}{2}\sqrt{1-\frac{r_0^2}{R^2}}\right)+\frac{R}{2}\arctan\left(\frac{r_0/R}{\sqrt{1-\frac{r_0^2}{R^2}}}\right)\right]\,.
\end{equation}
If $\Lambda\ll 1$ then
$$S(r_0)\approx\left(\frac{11V}{48\pi}\right)$$
which is proportional to the volume rather than the area as obtained
in the reference~\cite{PhysRevD.13.191} in the context of black
holes. It should be noted that, in this limit, $\Delta S\geq 0$,
since de Sitter Universe is in expansion.

Now let us allow the temperature $\Lambda$ to change. With such a
procedure we are interested on the specific heat at constant volume
$C_V$. Thus

$$C_V=\left(\frac{\partial E}{\partial T}\right)_V\,,$$
where keeping a constant volume means to have a constant $r_0$,
since $V=\frac{4\pi}{3}\,r_0^3$. Therefore we have

\begin{equation}
C_V=\frac{r_0^3}{3\,\sqrt{1-\frac{r_0^2}{R^2}}}\,,
\end{equation}
this quantity gives the information on how the gravitational energy
changes on the variation of the cosmological constant.

\section{Conclusion}
\noindent

In this article we have obtained the gravitational entropy for de
Sitter space-time in the framework of teleparallel gravity. Such a
result was obtained by purely thermodynamical quantities, i.e. using
concepts such as energy and pressure, that can be defined in TEGR,
we have derived an expression for the entropy. To obtain this we
have used the first law of Thermodynamics in which the cosmological
constant plays the role of the temperature. We have assumed that
because we have only two parameters in this system, the radius of
the hyper-surface of integration, $r_0$, which dictates how the
volume varies, and the cosmological constant, $\Lambda$. Thus the
temperature has to be a function of the cosmological constant, we
just have assumed the simplest form for it. We investigated the
entropy in the limit $\Lambda\ll 1$. We have obtained that the
entropy is proportional to the 3D volume, yet the entropy always
increases in this case, e.g. $\Delta S\geq 0$. Then we have relaxed
the condition $T=\Lambda=const.$ to obtain the specific heat at
constant volume, thus we derived the energy with respect to the
cosmological constant, keeping $r_0$ constant. If we have used the
same Hawking temperature, obtained  in the context of Schwarzschild
black hole, $T=1/2\pi$, then, in the limit $\Lambda\ll 1$, it would
neither be possible to find that the entropy is proportional to the
area nor to establish a specific heat, since the temperature is a
constant. Therefore the only possibility that generates an
expression for the entropy independent on the temperature, in such a
limit is $T=\alpha \Lambda$, where $\alpha$ is an arbitrary
constant.

%\bibliography{ref}
%\bibliographystyle{apsrev4-1}

\end{document}